\documentclass[twocolumn]{aastex631}

\usepackage{amsmath}
\usepackage{enumitem}

\begin{document}

\title{Multi-Resolution HEALPix Maps for Multi-Wavelength and Multi-Messenger Astronomy}

\correspondingauthor{Israel Martinez-Castellanos}
\email{imc@umd.edu}

\author[0000-0002-2471-8696]{I. Martinez-Castellanos}
\affiliation{Astroparticle Physics Laboratory, NASA Goddard Space Flight Center, Code 661, Greenbelt, MD 20771, USA}
\affiliation{Department of Astronomy, University of Maryland, College Park, MD 20742, USA}
\affiliation{Center for Research and Exploration in Space Science and Technology, NASA/GSFC, Greenbelt, MD 20771, USA}

\author[0000-0002-2471-8696]{Leo P. Singer}
\affiliation{Astroparticle Physics Laboratory, NASA Goddard Space Flight Center, Code 661, Greenbelt, MD 20771, USA}

\author[0000-0002-2942-3379]{E. Burns}
\affiliation{Department of Physics \& Astronomy, Louisiana State University, Baton Rouge, LA 70803, USA}

\author[0000-0002-9852-2469]{D. Tak}
\affiliation{Deutsches Elektronen-Synchrotron (DESY), Platanenallee 6, Zeuthen, 15738, Germany}

\author[0000-0001-5783-8590]{Alyson Joens}
\affiliation{George Washington University, 2121 I St NW, Washington, DC 20052, USA}
\affiliation{Astroparticle Physics Laboratory, NASA Goddard Space Flight Center, Code 661, Greenbelt, MD 20771, USA}

\author[0000-0002-4744-9898]{Judith L. Racusin}
\affiliation{Astroparticle Physics Laboratory, NASA Goddard Space Flight Center, Code 661, Greenbelt, MD 20771, USA}

\author[0000-0001-9608-4023]{Jeremy S. Perkins}
\affiliation{Astroparticle Physics Laboratory, NASA Goddard Space Flight Center, Code 661, Greenbelt, MD 20771, USA}

\begin{abstract}

HEALPix ---the Hierarchical Equal Area isoLatitude Pixelization--- has become a standard in high-energy and gravitational wave astronomy. Originally developed to improve the efficiency of all-sky Fourier analyses, it is now also utilized to share sky localization information. When used for this purpose the need for a homogeneous all-sky grid represents a limitation that hinders a broader community adoption.  This work presents \texttt{mhealpy}, a Python library able to create, handle and analyze multi-resolution maps, a solution to this problem. It supports efficient pixel querying, arithmetic operations between maps, adaptive mesh refinement, plotting and serialization into FITS ---Flexible Image Transport System--- files. This HEALPix extension makes it suitable to represent highly resolved region, resulting in a convenient common format to share spatial information for joint multi-wavelength and multi-messenger analyses.

\end{abstract}


\section{Introduction} \label{sec:intro}

HEALPix ---the Hierarchical Equal Area isoLatitude Pixelization \citep{HEALPix_Gorski_2005} \footnote{http://healpix.sourceforge.net}--- is a scheme to pixelate the sphere, originally developed to facilitate the analysis of the Cosmic Microwave Background (CMB) anisotropy. Its structural and geometrical properties allow for an efficient all-sky numerical Fourier decomposition and the convolution of local and global kernels. HEALPix has been implemented in multiple programming languages (C, C++, Fortran, IDL/GDL, Java, Python) and made available to the community since 1997.

HEALPix has grown beyond CMB analysis. Multiple astronomical fields have adopted it thanks to its convenient properties and accessible software libraries, for example, gamma-ray astronomy \citep{4FGL_Fermi_2020, Fourth-Fermi-GBM-2020, HAWC-Crab-2017}, high energy cosmic rays \citep{Auger_point_2014}, neutrino astronomy \citep{IceCube-7yrs-2017} and gravitational waves \citep{LIGO_bayestar_2016}. HEALPix is clearly established as a key tool for multi-messenger astronomy.

Having the astronomical data from various sources in the same format is enormously useful to communicate findings among the community and to perform joint analyses. A clear example is to report the sky localization of a source, especially when the uncertainty region is not well-described by a simple geometrical shape ---e.g. a circle or ellipse. Wide-spread adoption of a single format would not only make it easier to combine the localization information from various instruments, but it would also prompt collaborative efforts to develop  generic code --rather than instrument specific-- with the accompanying community support. Despite this, current limitations of the HEALPix standard and implementation have prevented a wider adoption. 

The main HEALPix limitation is that the data structure holds a full-sky map, at the same resolution everywhere, even when only a small portion is relevant (see \S\ref{sec:single-res-healpix}). While this was not a problem for its original use case, it severely restricts its use for multi-wavelength studies. For example, instruments with a high-precision localization usually find the amount of computer resources needed impractical or even unfeasible. 

The hierarchical structure of HEALPix provides a clear solution to this problem. Different levels of resolution can be realized for different regions by subdividing only certain pixels appropriately, as discussed in \S\ref{sec:multi-res-healpix}. This approach has been exploited either to generate a map efficiently, as a compression algorithm for storage or for faster querying  \citep{LIGO_bayestar_2016, MRH_Youngren_2017, ProgressiveSurveys_Fernique_2015}. It has also been utilized by the multi-order coverage (MOC) map IVOA (International Virtual Observatory Alliance) standard \citep{MOC_IVOA_2019} to specify arbitrary sky regions. The MOC standard was recently generalized by associating a value to each pixel of a full-coverage map, and is now used to distribute LIGO-Virgo gravitational wave sky localizations \footnote{LIGO/Virgo Public Alerts User Guide --- \url{https://emfollow.docs.ligo.org/userguide/tutorial/multiorder_skymaps.html}}.

Despite the fact that multi-resolution HEALPix maps are already being used by some collaborations, they are yet to have a broader adoption. This can be attributed to the lack of tools to perform the tasks frequently needed in multi-wavelengh and multi-messenger Astronomy. For example, the simple multiplication between two maps typically requires the rasterization of multi-resolution maps into a common single-resolution grid, defeating their purpose.

This work presents a new software package to generate and process muti-resolution HEALPix maps. It can create an adaptive multi-order mesh; query pixels and locations; efficiently perform binary operations between single or multi-resolution maps; and plot full or partial maps. Called \texttt{mhealpy}, it is an object-oriented extension of \texttt{healpy} \citep{healpy_Zonca_2019}, a Python wrapper for HEALPix C++. The code is available to the community and open-source\footnote{\url{https://mhealpy.readthedocs.io} \\(\url{https://doi.org/10.5281/zenodo.5706525})}.

In this document we first review the current HEALPix standard in \S\ref{sec:single-res-healpix}, followed by its generalization for multi-resolution maps in \S\ref{sec:multi-res-healpix}. Automatic ways to construct them are discussed in \S\ref{sec:adaptive-grids}, and the algorithms to query and operate on them in \S\ref{sec:operations}. Finally, some details of the implementation are presented on \S\ref{sec:implementation}.

\section{Single-resolution HEALPix maps}
\label{sec:single-res-healpix}

\begin{figure}
\centering
\includegraphics[width = \columnwidth]{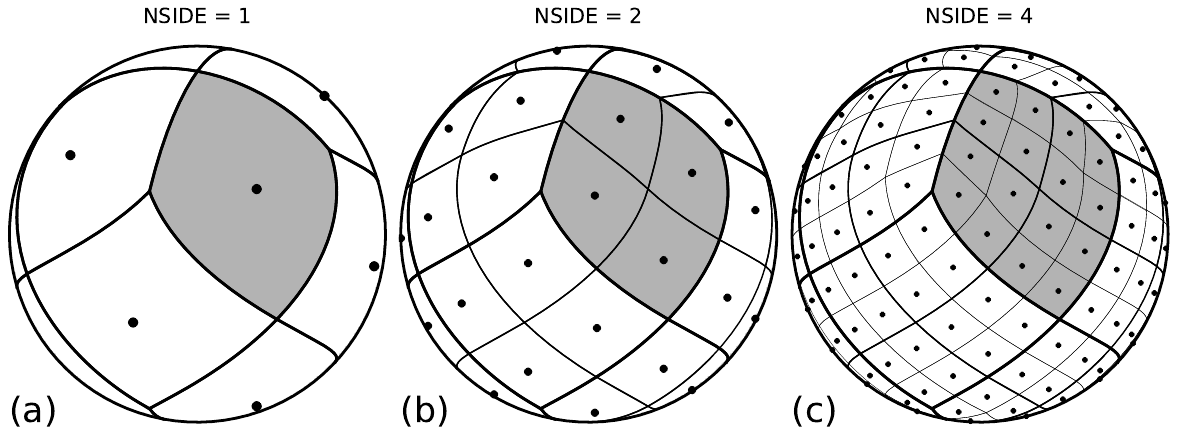}
\caption{Pixelization of the sphere for various \texttt{NSIDE} values. The center of the pixels, along iso-latitude lines, are represented with dots. A single base pixel is highlighted.}
\label{fig:grid_nside}
\end{figure}

\begin{figure}
\centering
\includegraphics[width = \columnwidth]{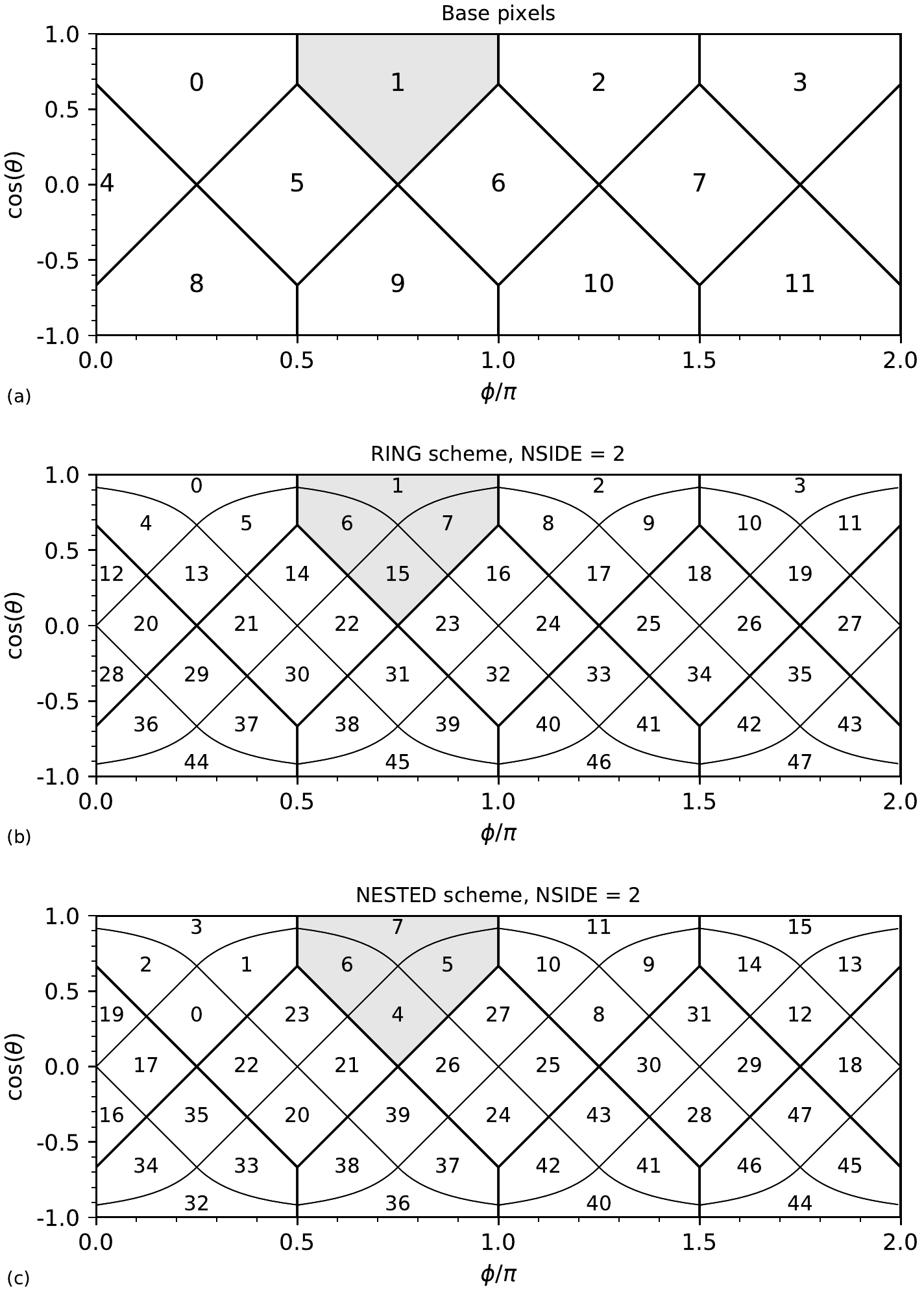}
\caption{a) Numbering of base HEALPix pixels ($\mathtt{NSIDE} = 1$). b,c) Pixel numbers for a map with $\mathtt{NSIDE} = 2$ under the \texttt{RING} and \texttt{NESTED} schemes, respectively.  A single base pixel is highlighted.}
\label{fig:grid_scheme_single}
\end{figure}

In the HEALPix standard the pixelization of the sphere begins by dividing the sphere into 12 base pixels of equal area, as shown in Figure \ref{fig:grid_nside}a and  Figure \ref{fig:grid_scheme_single}a. There are 4 base pixels around each pole and 4 around the equator. Each of these base pixels can be subdivided into 4 child pixels, and subsequently, each child pixel can be subdivided into 4 pixels of a lower hierarchy level. This process can continue until the desired resolution is achieved.

The number of subdivisions along a side of a base pixel is called \texttt{NSIDE} ---i.e. each base pixel is subdivided into $\mathtt{NSIDE}^2$ pixels. This is illustrated in Figure \ref{fig:grid_nside}. Alternatively, the pixel size can be specified using the resolution level $k$, also known as the map \textit{order}. This is an integer such that such that $\mathtt{NSIDE} = 2^k$.

The HEALPix standard specifies two different pixel numbering schemes. The \texttt{RING} scheme starts from zero at the North pole and increments from West to East and from North to South, along iso-latitude rings as shown in Figure \ref{fig:grid_scheme_single}b.  In the \texttt{NESTED} scheme the base pixels are labeled the same as in the \texttt{RING} scheme, but maps of higher \texttt{NSIDE} exploit the hierarchical nature of HEALPix to assign pixel numbers. As shown in Figure \ref{fig:grid_scheme_single}c the pixels are labeled such that the pixel $p$ in a map of order $k$ fully contains the pixels $[4p, 4(p+1))$ of a map of order $k+1$. These four child pixels are to be labeled in the following order: South, East, West and North.

\section{Geometry and labeling of multi-resolution maps}
\label{sec:multi-res-healpix}

\begin{figure}
\centering
\includegraphics[width = \columnwidth]{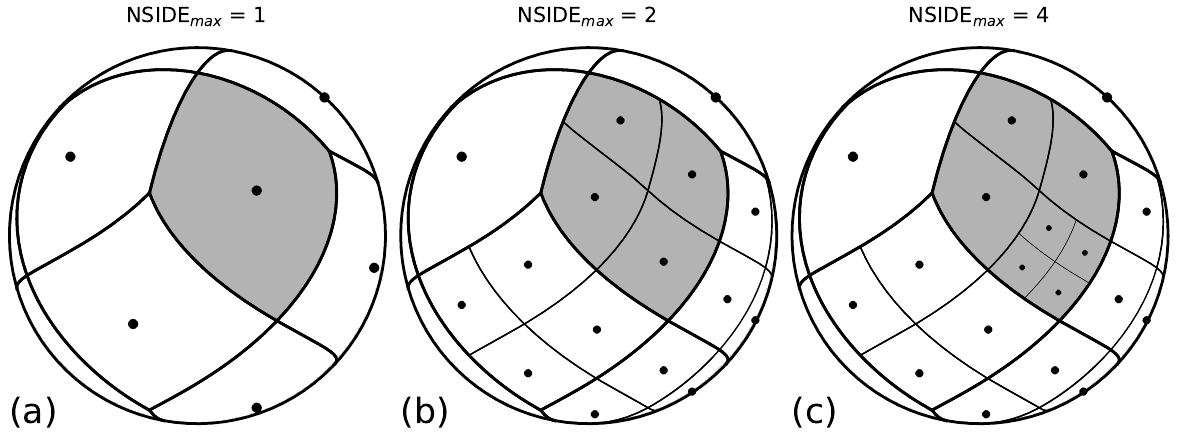}
\caption{Simple multi-resolution map. A single base pixel is highlighted (compare with Figure \ref{fig:grid_nside}). A cylindrical projection of (c) is shown in Figure \ref{fig:grid_scheme_multi}.}
\label{fig:grid_multi}
\end{figure}

Thanks to its hierarchical nature obtaining different resolution levels for different regions using HEALPix is straightforward. As shown in Figure \ref{fig:grid_multi} only a subset of the pixels can be subdivided with each iteration. Appropriately chosen, all pixels in an arbitrary region can be equal in size, or smaller, to the pixels in a map of an arbitrary \texttt{NSIDE}. 

In this work a multi-resolution map is treated as nothing more than a compression technique. The \texttt{NSIDE} of a map is then defined as the equivalent \texttt{NSIDE} for the smallest pixel it contains. Larger pixels are simply considered to represent a group of pixels that can be sufficiently well characterized by a single value. Consequently, user code can be mesh-agnostic, that is, no changes are needed to handle single or multi-resolutions maps.

\begin{figure}
\centering
\includegraphics[width = \columnwidth]{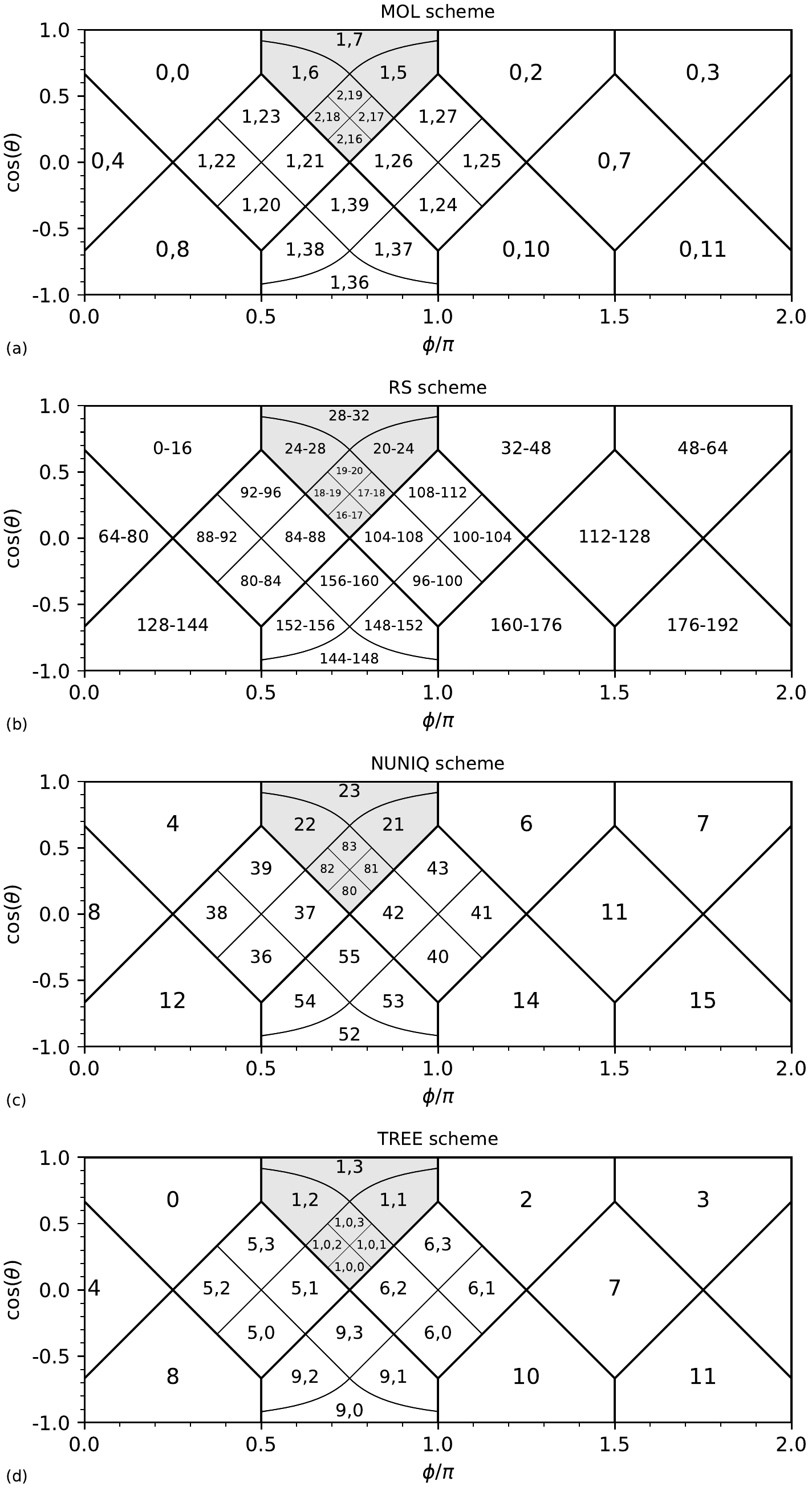}
\caption{Comparison of various labeling schemes for multi-resolutions maps. The figures are cylindrical projections of Figure \ref{fig:grid_multi}c. A single base pixel is highlighted. a) The \texttt{MOL} scheme labels each pixel by their equivalent order and pixel index in single-resolution \texttt{NESTED} maps. b) The \texttt{RS} scheme uses the beginning (inclusive) and end (exclusive) indices of a equivalent \texttt{NESTED} map with the highest resolution level. c) The \texttt{NUNIQ} encodes the equivalent order and nested pixel index into a single unambiguous number (explained in text). d) The \texttt{TREE} scheme stores the corresponding base pixel and the node indices in a quadtree.}
\label{fig:grid_scheme_multi}
\end{figure}

While there is consensus on how to construct a multi-resolution HEALPix map, there exists multiple possible ways to label the pixels. One possibility is to label each pixel using the number $p$ that would be assigned if it were a part of a single-resolution map (\texttt{NESTED} indexing is used here, although \texttt{RING} is also feasible). In addition, the equivalent order $k$ needs to be specified since different pixels from maps of different order share the same index. When arranged as one list of pixels per order this is called a multi-order list, or \texttt{MOL} scheme (see Figure \ref{fig:grid_scheme_multi}a). While this provides a compact encoding it is not well-suited for various algorithms ---e.g. arithmetic operations between maps, finding the pixel containing a given coordinate.

As suggested by \cite{Efficient_Reinecke_2015}, efficient algorithms can be developed if a given pixel is described by a range of equivalent pixels at the highest resolution level. Using an effective \texttt{NESTED} scheme, pixels are then labeled by their beginning (inclusive) and end (exclusive) index, that is, converting from the \texttt{MOL} scheme:
\begin{align}
rs_{start} &= 4^{k_{max}-k} p\\\nonumber
rs_{stop} &= 4^{k_{max}-k} \left(p+1\right)
\end{align}

This is referred as the range set (\texttt{RS}) scheme, and it is illustrated in Figure \ref{fig:grid_scheme_multi}b. Using the \texttt{RS} scheme it is also straightforward to validate that a map is well-formed. The condition for this is that every location in the sphere is contained by one and only one pixel. This is satisfied if, after sorting, the beginning of the first pixel equals 0, the beginning and end of all subsequent pairs match, and the end of the last pixel equals $12\,\mathtt{NSIDE}^2$.

For storage we adopt the nested unique (\texttt{NUNIQ}), proposed by \cite{Efficient_Reinecke_2015} and part of the IVOA recommendation for multi-order coverage maps \citep{MOC_IVOA_2019}. In this scheme the equivalent order and nested pixel number is encoded into a single unique number, suitable for serialization into a binary format. The \texttt{NUNIQ} numbering scheme (Figure \ref{fig:grid_scheme_multi}c) is defined as:
\begin{equation}
\mathtt{UNIQ} = 4^{k+1} + p
\end{equation}

Since $0 \leq p < 3\cdot4^{k+1}$ this single number identifies unambiguously any pixel in a map of any given \texttt{NSIDE}. The inverse operation is:
\begin{align}
k &= \left\lfloor \log_2\left( \mathtt{UNIQ}/4 \right)/2 \right\rfloor \\\nonumber
p &= \mathtt{UNIQ} - 4^{k+1}
\end{align}

Finally, during the generation of a multi-resolution HEALPix map, it is useful to label a pixel based on a series of indices $n_i$ that show the position of the pixel represented as a node in a tree structure, where $i$ runs from $0$ to $k$ (inclusive). The index $n_0$ is the base pixel number and the indices $n_{i>0}$ are within the range $[0,4)$ as seen Figure \ref{fig:grid_scheme_multi}d. We refer to this as the \texttt{TREE} scheme, as it explicitly recognizes each base pixels as a \textit{quadtree}. The equivalent pixel number in a \texttt{MOL} scheme is computed as:
\begin{align}
p = \sum_{i=0}^{k} 4^{k-i} n_i 
\end{align}

\section{Mesh generation}
\label{sec:adaptive-grids}

Different techniques can be used to efficiently generate a map with progressive refinements exploiting the hierarchical nature HEALPix --e.g. \cite{LIGO_bayestar_2016}. The output of these algorithms is a native multi-resolution map, which does not need to be rasterized into single-resolution maps in order to be distributed and analyzed. Multi-resolution maps, however, can be advantageous even if these methods are not used to compute a map.

An adaptive mesh refinement implementation is provided. This allows the user to automatically generate a mesh by increasing the resolution level of a given pixel using custom criteria. The algorithm accepts an arbitrary function $f(k,p)$ that decides whether a pixel should be split or added to the map as is. If split, the same evaluation is performed for the four pixels of order $k+1$ and indices $[4p,4(p+1))$. This recursion continues until there are no more pixels to split. This is implemented using the  \texttt{TREE} scheme in order to easily guarantee the resulting map has a valid mesh.

\begin{figure}
\centering
\includegraphics[width = \columnwidth]{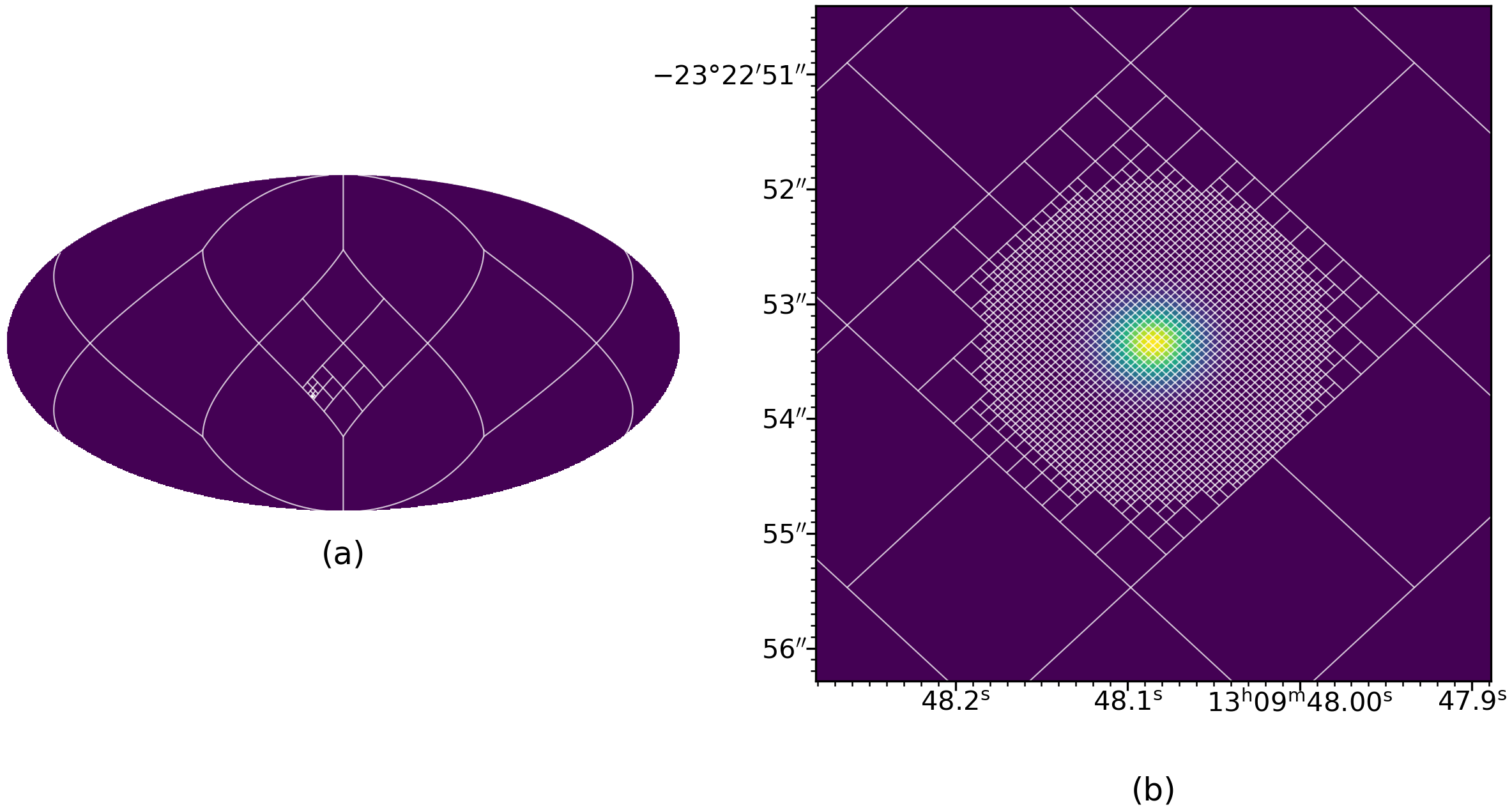}
\caption{A multi-resolution map representing the localization of SSS17a/AT 2017gfo by the Swope Telescope, both all-sky (a) and zoomed in around the source (b). The underlying mesh is overlaid. Sky maps for such well-localized sources are impractical or unfeasible using standard single-resolution maps.}
\label{fig:adaptive_SSS17a}
\end{figure}

Derived methods for common tasks are also provided. One of these occurs when we know \textit{a priori} the approximate region of the sky that requires high resolution and the rest of the sky simply needs to be filled  appropriately. This is exemplified in Figure \ref{fig:adaptive_SSS17a} with a map representing the localization of the optical transient SSS17a/AT 2017gfo by the Swope Telescope  \citep{SSS17a_Swope_2017}. A set of pixels $\left\{ p' \right\}$ contained in a circular region with a radius significantly larger than the localization error $\Delta\theta \approx 0.2''$ were selected from a single-resolution map of an appropriate order $k'$ such that the size of the pixels were a few times smaller than $\Delta\theta$. In order to have a valid map the rest of the pixels are selected by using the general adaptive mesh refinement method and choosing a splitting function $f(k,p)$ whose condition is whether there is any pixel number $p'$ within the range $[4^{k'-k}p, 4^{k'-k}(p+1))$. The SSS17a map of effective order $k = 22$ would be impractical or unfeasible to work with if it were a standard single-resolution map, but it is straightforward to handle using this approach.

\begin{figure}
\centering
\includegraphics[width = \columnwidth]{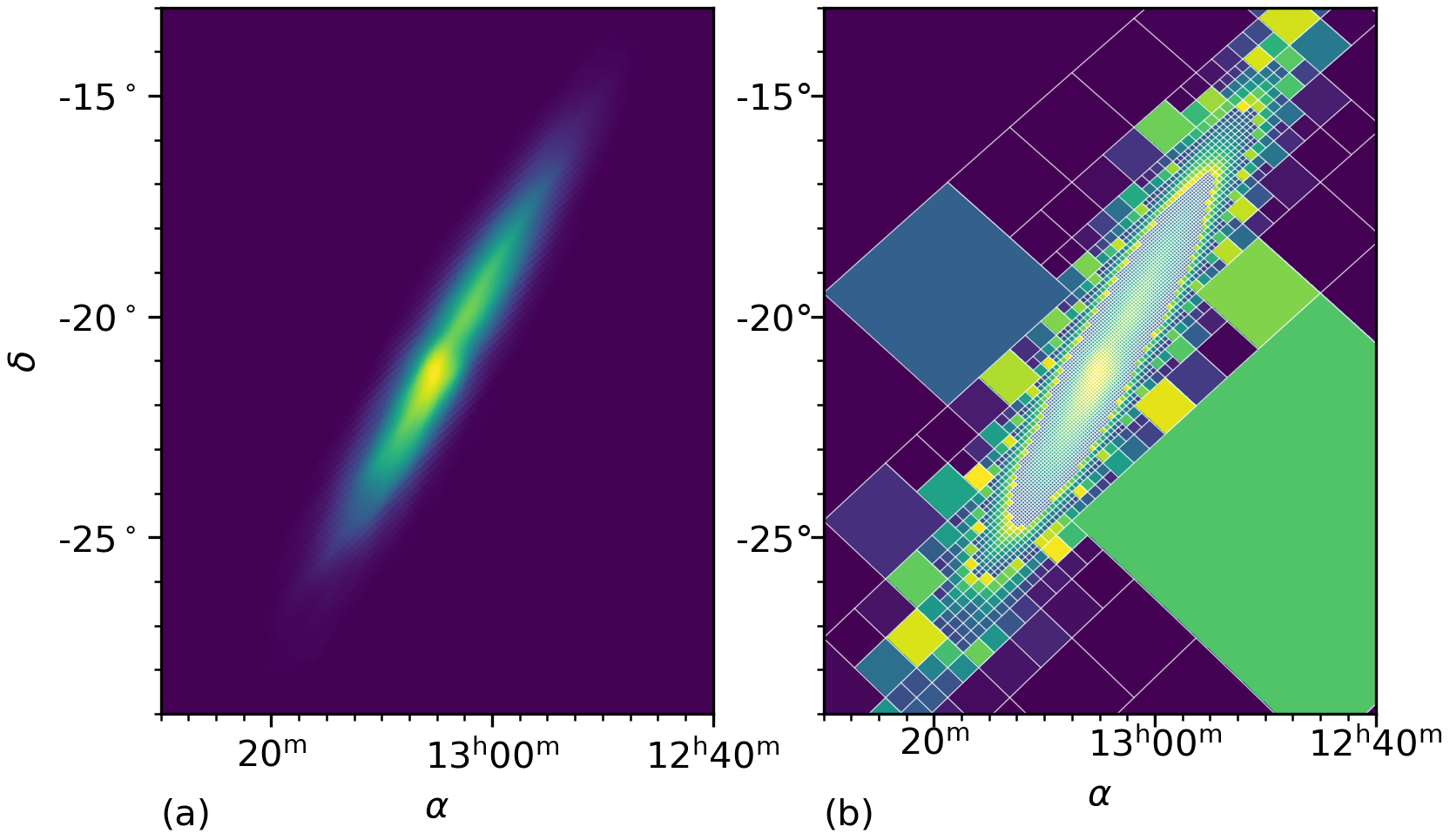}
\caption{a) GW 170817 sky localization probability density map by LIGO and Virgo \citep{GW170818-ligo-2017} b) Corresponding probability map integrated on the mesh resulting from an adaptive refinement process.}
\label{fig:adaptive_ligo}
\end{figure}

Another common task is to convert an existing single-resolution map into a multi-resolution map based on a maximum value threshold. In this case the pixels act as buckets with a maximum capacity and are split if this value is exceeded. For example, in Figure \ref{fig:adaptive_ligo} the GW 170817 sky localization probability map by LIGO and Virgo \citep{GW170818-ligo-2017} is compressed based on the condition that no pixel must contain a greater probability that the maximum value of the original single-resolution map of order $k'$. That is, the criterion for the splitting function  $f(k,p)$ is whether the sum of all pixels $p_i$ within the range $[4^{k'-k}p, 4^{k'-k}(p+1))$ is greater than the threshold value. In the output multi-resolution map all pixels carry a comparable weight, and the sampling resolution is proportional to the integrated probability contained in a given region. This algorithm is a form of quadtree decomposition frequently used for image compression \citep{quadtree_Shusterman_1994}.  

\section{Map operations}
\label{sec:operations}

Map operations can be either related to the mesh itself, in the following referred to as ``pixelization operations'', or binary arithmetic operations between maps ---e.g. addition, multiplication. The following section explains how these are implemented for multi-resolution maps.

\subsection{Pixelization operations}

Common pixel-related operations are pixel to coordinates conversion ---i.e. computing the center and boundaries of a pixel---, digitization ---i.e. obtaining the pixel a given coordinate belongs to---, and pixel querying ---i.e finding the pixels contained in or overlapping with a given region, for example a disc. Most of these were generalized to multi-resolution maps using routines written for standard single-resolution maps by choosing an appropriate labeling scheme. In particular, converting from a pixel to a coordinate is trivial in the \texttt{MOL} scheme.

The digitization is performed efficiently using the \texttt{RS} scheme. First the equivalent pixel number in a \texttt{NESTED} map of the same order is obtained, followed by a search for the actual pixel whose range contains this number. Presorting pixels based on their \texttt{RS} representation makes this and other procedures more efficient.

A similar strategy was used for pixel querying. All equivalent \texttt{NESTED} pixels overlapping  a given region are found and then a search in a sorted \texttt{RS} list is performed. This works well when querying for overlapping pixels, since any pixels with at least one child pixel overlapping a region will also be itself an overlapping pixel. However, sometimes the user queries all pixels whose centers lie within a region. The previous strategy fails here since the center of a child pixel might be contained inside the region of interest while the center of the parent lies outside, even it they partially overlap. In this case, a robust albeit less efficient solution is to use the \texttt{MOL} representation and search order by order. 

\subsection{Binary arithmetic operations}

Operations between standard single-resolution maps are performed pixel by pixel. If both operands have the same  \texttt{NSIDE} and numbering scheme this is a straightforward and unambiguous operation. When that's not the case, it is left to the user to either upgrade or downgrade the resolution of one of the maps, and swap the scheme if needed. 

Various interpolation algorithm can be used to increase the resolution of a map. At zeroth order however, and for the purpose of a binary operation, it is enough to divide each pixel of the coarser map into $4^{\Delta k}$ child pixels of equal value. This value depends on whether the quantity in a map depends on the solid angle area covered by a pixel --e.g counts, probability or any extensive property-- or not --e.g. temperature, probability density or any intensive property. That is, the child pixels in a density-like map are assigned the same value as their parent, while in a histogram-like map the contents of the parent pixel are split equal ways among the children. The inverse process occurs when downgrading the resolution.

\begin{figure}
\centering
\includegraphics[width = \columnwidth]{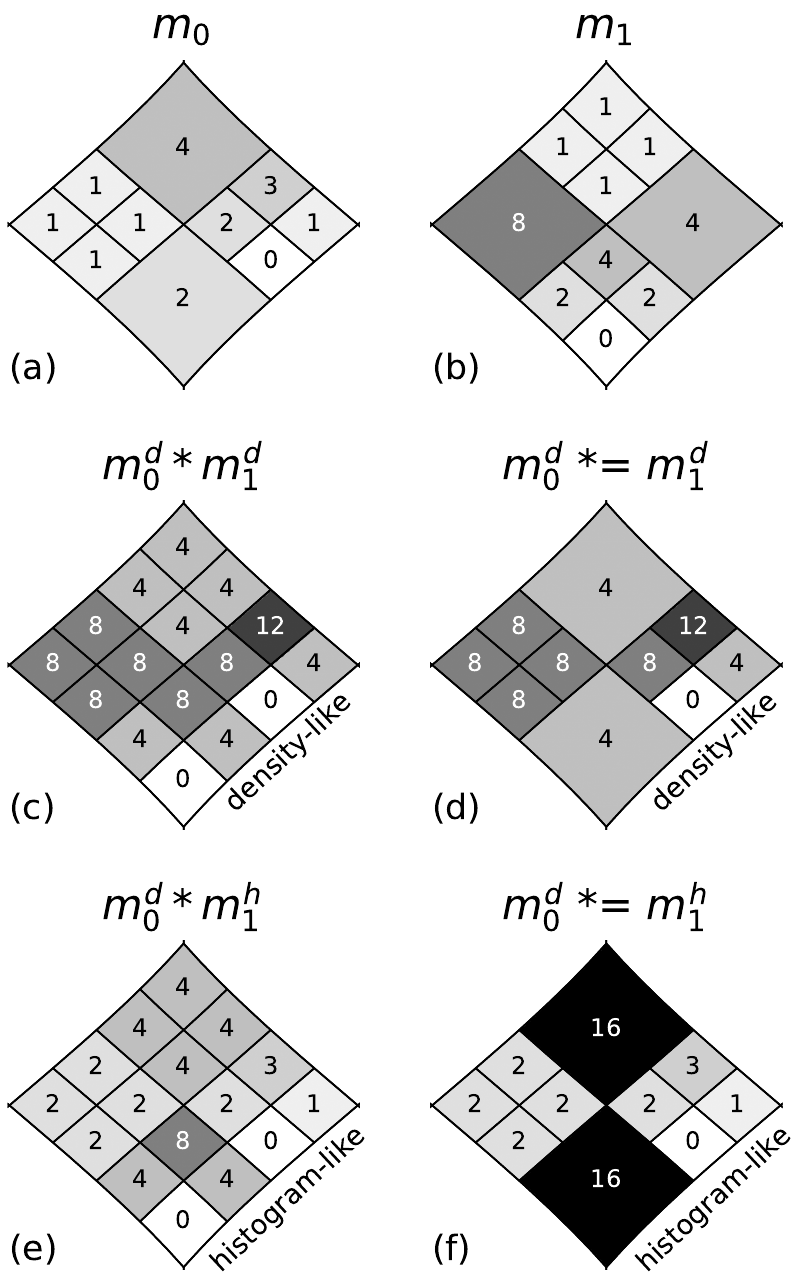}
\caption{Example of the result of taking the product between two maps, (a) and (b), depending on various conditions. A single base pixel is shown. Figures (c) and (d) assume both maps are density-like, resulting also in a density-like map. Figures (e) and (f) consider $m_0$ to be density-like and $m_1$ histogram-like, the output therefore being histogram-like. Standard operations, with no information loss, are shown in (c) and (e), while in-place operations that keep the $m_0$ mesh unchanged result in (d) and (f).}
\label{fig:operation_product}
\end{figure}

Binary operations between multi-resolution maps follow this same approach. During this process the algorithm combines or split pixels as needed, assigning appropriate values to the child pixels depending on whether the map is histogram-like or density-like as defined by a flag set by the user. Compare Figures \ref{fig:operation_product}c and \ref{fig:operation_product}d to Figures \ref{fig:operation_product}e and \ref{fig:operation_product}f. 

Any operation between two density-like maps results in a density-like map. Addition or subtraction between histogram-like maps results also in a histogram-like map. The product of a density-like map with a histogram-like map is histogram-like. The ratio between two histogram-like maps is a density-like map. The result from all other combinations defaults to the same type as the leftmost operand. Note however that these are in general ill-defined for physical quantities and can result in mesh-dependent results. Operations with scalars leave the map type unchanged.

Binary operations can be performed efficiently by having both maps in a sorted \texttt{RS} representation scheme, with the pixel ranges in both lists corresponding to the same \texttt{NSIDE}, the largest of the two maps. The algorithm then proceeds in order splitting or combining pixels as needed in order to match the mesh between the two operands. This is a general approach that can accept any two maps as input, whether they are single or multi-resolutions maps, and whether they do or do not have the same resolution or numbering scheme.

We distinguish between standard and in-place operations. During standard binary operations, the pixel of the coarser map within a given equivalent pixel range is always split to match the mesh of the map with the highest resolution in that region. This ensures that there is no loss of information, as shown in Figures \ref{fig:operation_product}c and \ref{fig:operation_product}e.

During an in-place operation the pixels of the second operand are either split or combined appropriately to match the pixels on the first map. The mesh of this map remains the same and the pixel values are updated on the go. No additional memory needs to be allocated, at the expense of potential information loss. This is exemplified in Figures \ref{fig:operation_product}d and \ref{fig:operation_product}f. In-place operations can also be used to rasterize a multi-resolution map by taking the product with a single-resolution map whose pixels have all been initialized to one.

\section{Implementation}
\label{sec:implementation}

The ideas presented in \S\ref{sec:adaptive-grids} and \S\ref{sec:operations} were implemented in the Python library \texttt{mhealpy}. The only direct dependency is the \texttt{healpy} library,  which in turns is a wrapper to the HEALPix C++ library. Similar nomenclature makes it easy to refactor code using \texttt{healpy} and generalize it to handle multi-resolution maps.

The \texttt{mhealpy} library uses an object-oriented design. The user does not need to keep track of the various map properties. Similar to the C++ implementation, most of the code is either accessed through the \texttt{HealpixBase} or the \texttt{HealpixMap} classes. The former stores the mesh and numbering scheme, and contains all the pixelization operations which do not involve the map contents.  \texttt{HealpixMap} is a derived class of \texttt{HealpixBase} which in addition contains the map contents and whether it is a histogram-like or density-like map.

Most operations available in \texttt{healpy} were generalized in \texttt{mhealpy} for their use with multi-resolution maps without rasterizing them into single-resolution maps first, which saves computer resources. This includes plotting. Spherical harmonic transforms are an exception, they were not implemented for multi-resolution maps since it is more efficient to work with standard HEALPix maps.

The map serialization is complaint with the  IVOA standard \citep{MOC_IVOA_2019} for multi-order coverage maps. Maps are saved into FITS \citep{FITS-Hanish-2001} tables with an explicit \texttt{NUNIQ} ordering. This is readily compatible with LIGO-Virgo skymaps which adopted the same format.

\section{Summary}

This work introduced \texttt{mhealpy}, an object-oriented wrapper of the \texttt{healpy} Python library that allows to create and handle multi-resolution HEALPix maps. This extension to the current standard prevents the need for a homogeneous all-sky grid, making it a viable option as a common format to share spatial information for joint multi-wavelength and multi-messenger analyses, including for well-localized detections. We presented the definition of multi-resolution maps, various pixel labeling schemes, and how the various functionalities were implemented: pixel querying, binary arithmetic operations, adaptive grid refinement, plotting and serialization into FITS files.

\begin{acknowledgments}
We would like to thank Boyan A. Hristov (UAH) for the valuable comments that improved this work significantly. The material is based upon work supported by NASA under award number 80GSFC21M0002. Some of the results in this paper have been derived using the healpy and HEALPix package.
\end{acknowledgments}

\bibliography{ref}{}
\bibliographystyle{aasjournal}

\end{document}